\documentclass[amsmath,amssymb,aps,prl,twocolumn]{revtex4}
\usepackage{graphicx}
\usepackage{array}
\usepackage{amsmath,amsfonts,amssymb}
\usepackage[ansinew]{inputenc}
\usepackage{color}
\usepackage{calc}
\usepackage[normalem]{ulem}

\usepackage{graphicx}% Include figure files
\usepackage{dcolumn}% Align table columns on decimal point
\usepackage{bm}% bold math
\newcommand{\Figref}[1]{Fig.~\ref{#1}}

\begin{document}

\title{Many-body description of STM-induced fluorescence of charged molecules}
\author
{Song Jiang$^{1\ast}$, Tom{\'a}\v{s} Neuman$^{2\ast}$, R\'emi Bretel${^2}$, Alex Boeglin$^1$, \\
 Fabrice Scheurer$^1$, Eric Le Moal$^{2}$, Guillaume Schull$^{1\ast}$\\
\normalsize{$^1$ Universit\'e de Strasbourg, CNRS, IPCMS, UMR 7504, F-67000 Strasbourg, France,} \\
\normalsize{$^2$ Institut des Sciences Mol\'eculaires d'Orsay (ISMO), UMR 8214, CNRS, Universit\'e Paris-Saclay, 91405 Orsay Cedex, France.}\\
\altaffiliation{song.jiang@ipcms.unistra.fr; tomas.neuman@universite-paris-saclay.fr; guillaume.schull@ipcms.unistra.fr}
}%

\date{\today}

\begin{abstract}
A scanning tunneling microscope is used to study the fluorescence of a model charged molecule (quinacridone) adsorbed on a sodium chloride (NaCl)-covered metallic sample. Fluorescence  from the neutral and positively charged species is reported and imaged using hyper-resolved fluorescence microscopy. A many-body excitation model is established based on a detailed analysis of voltage, current and spatial dependencies of the fluorescence and electron transport features. This model reveals that quinacridone adopts a large palette of charge states, transient or not, depending on the voltage used and the nature of the underlying substrate. 
%Overall, the universal character of the proposed model shed light on transport and luminescence mechanisms of molecules adsorbed on thin insulators.
This model has a universal character and explains the electronic and fluorescence properties of many other molecules adsorbed on thin insulators.

\end{abstract}
%revealing transition dipole moments oriented at $\approx$ 65$^{\circ}$ from each others.

\maketitle

Fluorescence of neutral and charged molecules has been reported with sub-nanometer resolution in scanning tunneling microscopy induced-luminescence (STML) experiments over the last few years \cite{qiu2003vibrationally,zhang2016visualizing,imada2016real,Doppagne2017Vibronic,doppagne2018electrofluorochromism,kimura2019selective,chen2019spin,Kaiser2019,Rai2020,kong2021,Hung2021,dolezal2021exciton,dolezal2021real}. In these experiments, charged emission is generally probed during transient charging of neutral molecules adsorbed on a thin decoupling layer \cite{doppagne2018electrofluorochromism,Rai2020,dolezal2021exciton}. STML of molecules whose ground state on surface is a negatively charged doublet (D$_0^{-}$) has also been reported for perylenetetracarboxylic dianhydride (PTCDA) in two recent reports \cite{kimura2019selective,dolezal2021real}. In these experiments, a low energy emission line was reported, which was assigned either to phosphorescence or to the emission of the negatively charged molecule (negative trion). For PTCDA, deciphering which of those two emission channel is at play is tedious, especially because the transition dipole moments of both phosphorescence and negative trions are oriented along the same molecular axis and would give rise to similar patterns in STML maps. 

Here, we report on a STML study of a model quinacridone (QA) molecule, whose ground state is a positively charged doublet (D$_0^{+}$) when adsorbed on 4ML NaCl/Ag(111). A rare property of this molecule is that neutral and positively charged fluorescence dipoles are oriented at $\approx$ 65$^{\circ}$ from each other, giving rise to radically different patterns in STML maps. This allows us to undoubtedly exclude phosphorescence processes and to assign the low energy emission line to the positive QA trion. Based on a comparison between the spatial, voltage and current dependencies of the charged and neutral emission lines, and on conductance spectra, an excitation mechanism based on a many-body description of the system \cite{schulz2015many, Yu2017, miwa2019,doppagne2020single,fatayer2021probing,song2022} can be set, which reveals that QA can be populated in four different charged states (namely QA$^-$, QA$^0$, QA$^+$, and QA$^{2+}$) and three different spin multiplicity (singlet, doublet, triplet) during a single voltage sweep. This excitation mechanism is backed up by data obtained on the same QA molecule deposited on NaCl on Au(111) where the ground state of the molecule is a doubly positively charged singlet state (S$_0^{2+}$). We believe that this excitation model bears a universal character that can be applied to STML experiments of both charged or neutral molecules.

The STM data were acquired with a low temperature (4.5~K) Omicron setup operating in ultrahigh vacuum adapted to detect the light emitted at the tip-sample junction (see Supplemental Material for details). The Ag(111) or Au(111) substrate was cleaned with successive sputtering and annealing cycles. Electrochemically etched gold tips were gently introduced in the substrate to tune their plasmonic response. NaCl was thermally evaporated on substrate maintained at room temperature. The sample was then annealed to  $\sim$ 400-430~K to obtain NaCl films with thicknesses of three to four monolayers(ML). Eventually, the NaCl-covered substrate was introduced into the STM chamber and cooled down to $\sim$ 4.5~K. QA molecules were thermally sublimated (at $\sim$ 560~K) from a quartz crucible onto the cold substrate. 

\begin{figure}[ht]
\includegraphics[width=8.5cm]{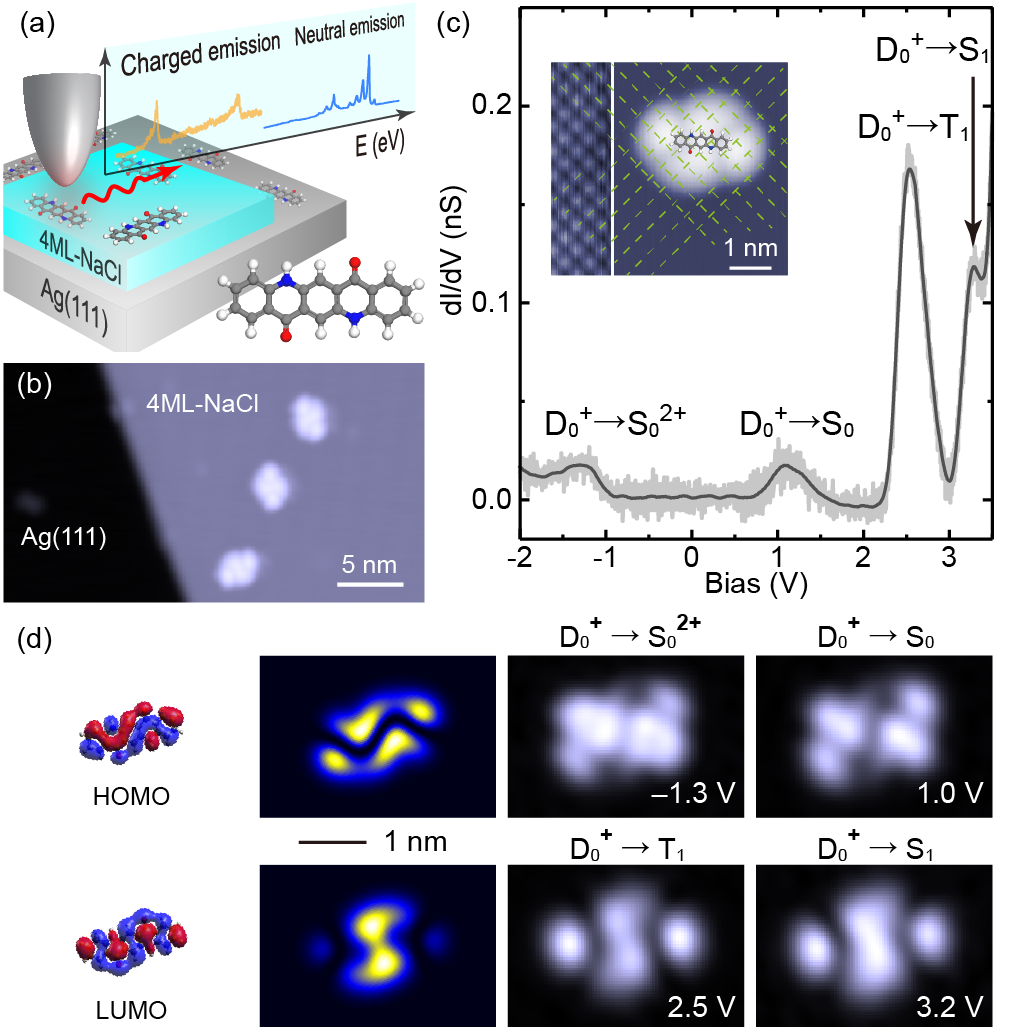}
\caption{\label{fig1} %\textbf{Electronic structures of QA molecules.}
(a) Sketch of the STML experiment on NaCl-decoupled QA molecules where neutral and charged fluorescence are represented. (b) Typical STM image ($V$= 2.8 V, $I$ = 5 pA) of QA molecules adsorbed on 4ML-NaCl islands on Ag(111). (c) d$I$/d$V$ spectrum recorded at the center of the QA molecule on 4ML-NaCl. Inset: molecular adsorption configuration (the left part is acquired at $V$ = 0.4 V and $I$ = 500 pA , and the right part is acquired at $V$ = 2.6 V and $I$ = 5 pA). (d) DFT calculations (see Supplementary Materiel for details) of the frontier orbitals of QA molecule and the corresponding maps of their partial density of states at constant height are shown alongside with experimental d$I$/d$V$ maps acquired at voltages corresponding to d$I$/d$V$ resonances in (c).} 
\end{figure} 

Figure \ref{fig1}a shows a sketch of the experiment, where the tip of a scanning tunneling microscope (STM) is used to excite the fluorescence QA on 4 ML NaCl/Ag(111). In \Figref{fig1}b one identifies individual QA molecules adsorbed on the NaCl decoupling layer.
Figure \ref{fig1}c provides a typical conductance spectrum (d$I$/d$V$) acquired on the decoupled QA, together with a close-up view of a QA molecule revealing its detailed adsorption site on NaCl. Here, the  two oxygen atoms of the molecule are located above a Sodium atom. 
%Eventually, constant height d$I$/d$V$ maps recorded for each of the resonances observed in \Figref{fig1}c are displayed in \Figref{fig1}d together with  density functional theory (DFT) representations of the  highest occupied molecular orbital (HOMO) and the lowest unoccupied molecular orbital (LUMO) of the neutral QA molecule. 
Interestingly, the d$I$/d$V$ spectrum reveals two nearly symmetric resonances below ($V$ = $-$1.3 V) and above ($V$ = 1.0 V) the Fermi level, whose associated maps (\Figref{fig1}d) reveal patterns strongly resembling the shape of the DFT-calculated HOMO (highest occupied molecular orbital). Indeed, in this electronic configuration, adding (D$_0^{+} \rightarrow$ S$_0$) or removing (D$_0^{+} \rightarrow$ S$_0^{2+}$) an electron from the molecule with the tip involves the HOMO of the QA molecule. Similar conclusions have been reported in previous STM works one decoupled molecules \cite{repp2006imaging,hollerer2017charge,kimura2019selective}; they generally indicate that the molecular ground state on NaCl is the positively charged doublet state (D$_0^{+}$).  

In contrast, the d$I$/d$V$ resonances recorded at higher positive voltages reveal patterns that are characteristic of the LUMO (lowest unoccupied molecular orbital). For reasons that will become clear later, we assign these resonances to transitions to the triplet (D$_0^{+} \rightarrow$ T$_1$) and excited singlet (D$_0^{+} \rightarrow$ S$_1$) state of the neutral molecule. The resonance corresponding to a transition to the triplet state appears notably more intense than the transition to the excited singlet state, a phenomenon that can be  associated to the larger spin multiplicity of this state (See Supplemental Material). The role of molecular triplet states was inferred or more directly measured in several scanning probe measurements recently \cite{kimura2019selective, chen2019spin, fatayer2021probing, peng2021}. Identifying it here as a resonance in d$I$/d$V$ spectra, and clarifying its role in emission spectra is therefore of importance.     
%\sout{To our knowledge, it is the first time that a resonance in a d$I$/d$V$ spectrum is directly associated to a triplet state.} 

\begin{figure}[t]
\includegraphics[width=8.5cm]{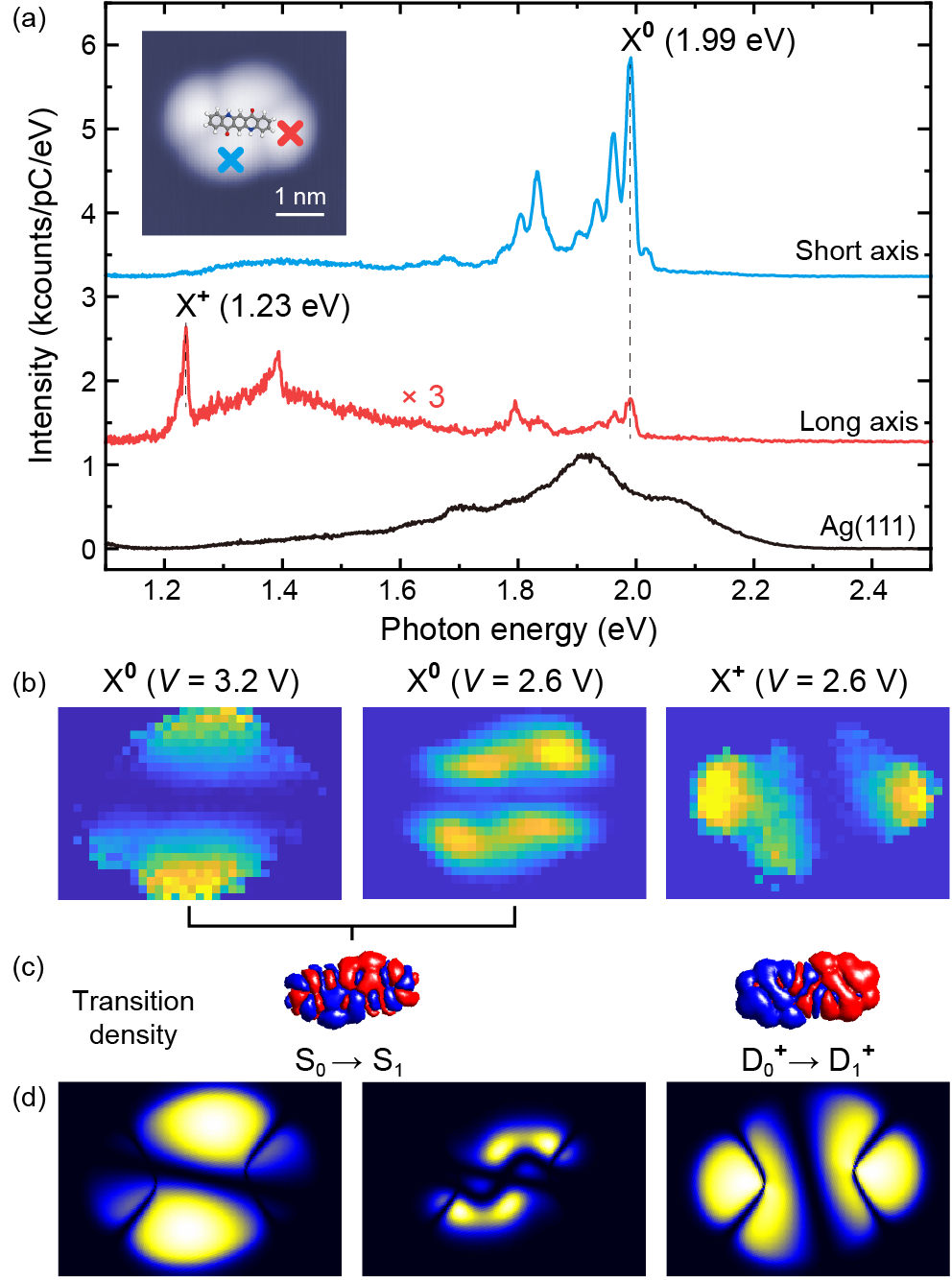}
\caption{\label{fig2} %\textbf{Neutral and charged emission from individual QA molecules.}
(a) STML spectra acquired ($V$= 2.7 V, $I$ = 100 pA, acquisition time t = 300 s) for the STM tip located at the positions identified in the inset by crosses ($V$= 2.6 V, $I$ = 5 pA). The spectrum acquired on Ag(111) is measured with ($V$= 2.7 V, $I$ = 200 pA, $t$ = 30 s). (b) HRFMs of QA on NaCl for the neutral (X$^0$) and charged emission (X$^+$). Mapping conditions: 3.5 nm $\times$ 2.4 nm, 35 $\times$ 24 pixels, $V$ = 2.6 V, $t$ = 30 s per pixel.  (c) Transition electron density associated to the S$_0$ $\rightarrow$ S$_1$ and D$_0^+$  $\rightarrow$ D$_1^+$ transitions calculated using TDDFT. (d) Simulations of the HRFMs (see text for details).} 
\end{figure}

In  \Figref{fig2}a we show STM-induced luminescence (STML) spectra acquired over the short (in blue) and long (in red) axes of QA at a bias of $V$ = 2.7 V, together with a reference spectrum acquired on the bare Ag(111) (in black). The blue spectrum reveals a high energy emission line at h$\nu$ = 1.99 eV (labeled X$^{0}$) followed by a set of vibronic lines at slightly lower energy that fit well the emission of neutral QA (S$_1 \rightarrow$ S$_0$ transition). The same contribution can be identified in the red spectrum, but with a much weaker emission intensity. It is accompanied by two low energy lines at h$\nu$ = 1.23 eV (labeled X$^{+}$) and h$\nu$ = 1.39 eV, whose identification cannot rely only on literature.

In  \Figref{fig2}b we display hyper resolved fluorescence maps (HRFMs) of the emission intensity of the X$^{0}$ line (recorded at $V$ = 3.2 V and $V$ = 2.6 V) and of the X$^{+}$ line (recorded at $V$ = 2.6 eV). 
%Acquired at a bias of $V$ = 2.6 V, the current map reveals the expected LUMO pattern.
The HRFMs exhibit two bright features separated by a nearly horizontal (X$^{0}$ line) or vertical (X$^{+}$ line) node. It has been demonstrated in earlier works \cite{chen2010, zhang2016visualizing, doppagne2020single,roslawska2022prx}, that the pattern in HRFMs is characteristic of the excited transition dipole moment and its orientation. More precisely, the axis joining the two bright lobes in the HRFMs provides the dipole orientation. One notes that for the X$^{0}$ line, the dipole axis is tilted by $\approx 65^\circ$ with respect to the long axis of QA,
%an expected behaviour for the S$_1 \rightarrow$ S$_0$ transition. This is confirmed by our TDDFT calculations performed in Gaussian 16 on the B3LYP/6-31G(d,p) level of theory (see the Supp. Mat. for details). The simulated transition densities appear in \Figref{fig2}c.
as shown by the TDDFT-simulated  S$_1 \rightarrow$ S$_0$ transition dipole in \Figref{fig2}c (see Supplemental Material for details), that is a transition of the neutral molecule.
Note that the same orientation is expected for the dipole associated to the phosphorescence T$_1 \rightarrow$ S$_0$ of the QA molecule. In contrast, the HRFM of the h$\nu$ = 1.23 eV line reveals a transition dipole moment nearly parallel to the long QA axis. This behaviour agrees very well with TDDFT simulations (\Figref{fig2}c) of the positively charged QA trion (X$^{+}$) corresponding to a D$_1^{+} \rightarrow$ D$_0^{+}$ radiative transition. For the 1.39 eV line, the HRFM reveals a pattern similar to the one for X$^{+}$ (see Supplemental Material). Together with spectra recorded in the short-wave infrared range (see Supplemental Material) that confirm the assignment of the 1.23 eV line to the 0-0 of X$^{+}$, these data associate the 1.39 eV line to the luminescence hot trions (\textit{i.e.} emission from a non-relaxed vibronic state of D$_1^{+}$).  Eventually, the X$^0$ HRFMs recorded at $V$ = 3.2 V and $V$ = 2.6 eV reveal two slightly different patterns, a difference that reflects different excitation mechanisms as can be accounted for in a more complex theoretical model, as described below (see also \Figref{fig2}d).
%This model, described in detail below, allows us to reproduce the experimental HRFM maps with an excellent agreement (the theoretically calculated maps are shown in \Figref{fig2}d). 

\begin{figure}[t]
\includegraphics[width=8.5cm]{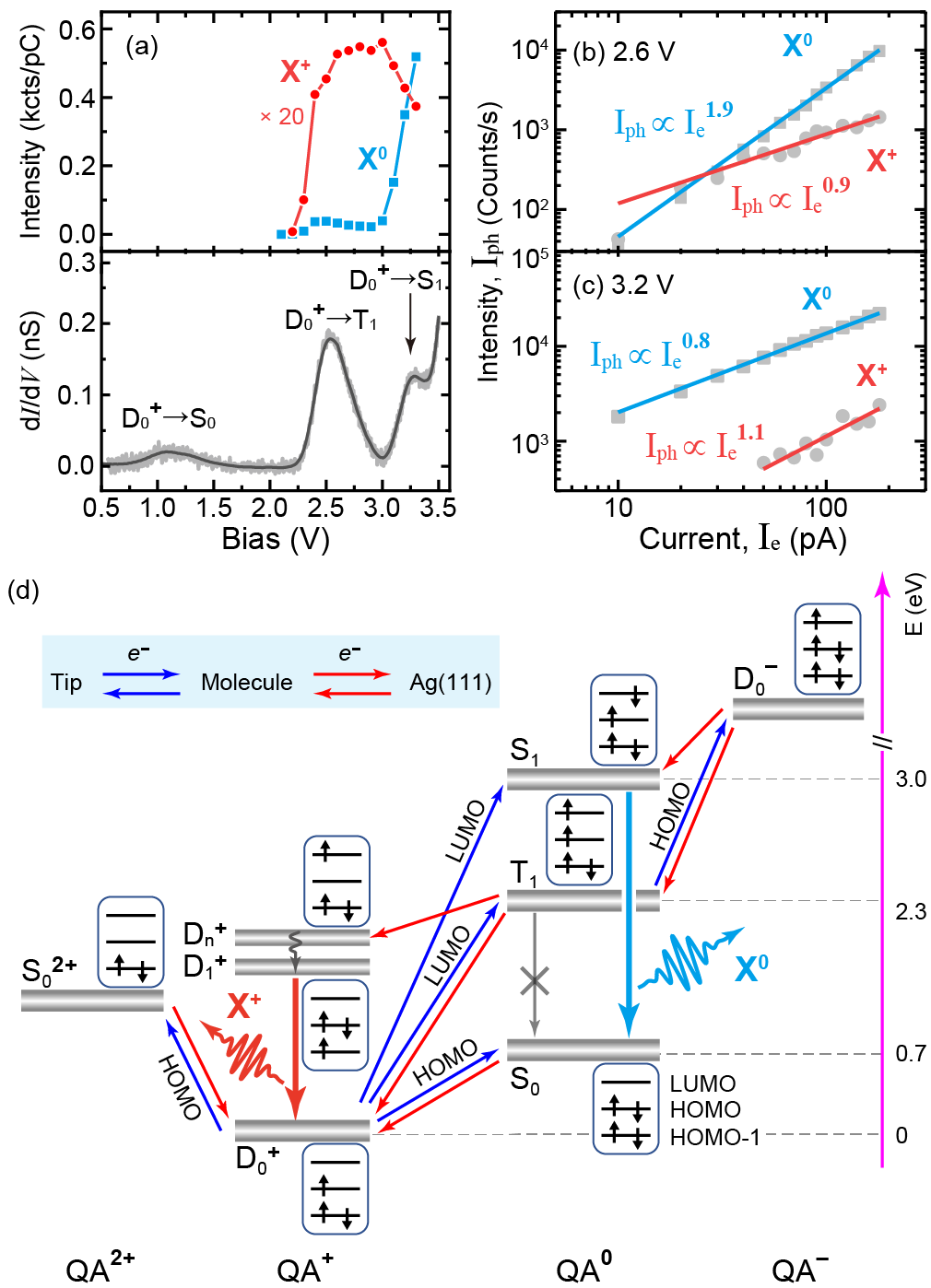}
\caption{\label{fig3} %\textbf{Luminescence mechanisms for the neutral emission  and charged emission from QA molecules.}
(a) Voltage dependencies of the neutral (X$^0$, blue squares) and charged (X$^+$, red circles)  emission intensities measured at a constant current of $I$ = 50 pA. A ${\rm d}I/{\rm d}V$ spectrum of a QA/NaCl(4ML)/Ag(111) is shown as a reference (bottom panel). (b and c) Current dependencies of the of neutral (X$^0$, blue squares) and charged (X$^+$, red circles) emission intensities measured at 2.6 V (b) and 3.2 V (c). (d) Sketches of the luminescence mechanisms for the neutral and charged QA emission.} \end{figure}

To support our assignment of the spectral contributions and to determine with precision the excitation mechanism leading to the emission of both charged and neutral QA molecules, we recorded the voltage (\Figref{fig3}a) and current (\Figref{fig3}b,c) dependencies of the X$^{0}$ and X$^{+}$ lines. Figure \ref{fig3}a shows that the emission onset of X$^{+}$ coincides with the resonant tunneling into the triplet state (D$_0^{+} \rightarrow$ T$_1$). Besides, and independently of the used voltage bias, the X$^{+}$ emission intensity is linear with current (\Figref{fig3}b,c), indicating a single electron excitation process. Based on these observations we can build a robust many-body scheme (\Figref{fig3}d) to explain the X$^{+}$ emission mechanism.  The molecule initially in the ground D$_0^{+}$ state can be driven, for V $>$ 2.3 V, into T$_1$ by injection of an electron from the tip directly into the LUMO of the molecule. The molecule can turn back to its cationic state by electron tunneling to the substrate either from the LUMO (T$_1 \rightarrow$ D$_0^{+}$) or from the HOMO, populating thus a higher excited state D$_n^+$ (T$_1 \rightarrow$ D$_n^{+}$)\cite{note}. In the latter case, the trionic excited state D$_n^{+}$ rapidly decays to D$_1^{+}$. The radiative transition (D$_1^{+} \rightarrow$ D$_0^{+}$) is eventually responsible for the  X$^{+}$ emission line (as confirmed by TDDFT, see Supplemental Material). The only condition for this mechanism to occur is that T$_1$ has a higher energy than D$_n^{+}$ in the diagram of \Figref{fig3}d.

The situation is more complex for the X$^{0}$ emission. The voltage dependency in \Figref{fig3}a reveals here two emission onsets: A first one that leads to a weak X$^{0}$ emission and that coincides with the (D$_0^{+} \rightarrow$ T$_1$) transition in the d$I$/d$V$ spectrum, and a second onset leading to a much brighter emission intensity at a voltage that matches the (D$_0^{+} \rightarrow$ S$_1$) transition. The plot of \Figref{fig3}b reveals a nearly quadratic dependency of the X$^{0}$ intensity with current for voltages corresponding to the first onset. This last observation clearly indicates a two-electron excitation mechanism. In contrast, the linear dependency of the X$^{0}$ emission after the second onset (\Figref{fig3}c) indicates a one-electron process for this voltage range. 

One easily understands the observation of an intense X$^{0}$ emission for V $>$ 3.0 V: above this voltage, one can populate the singlet excited state S$_1$ of the molecule, which in turn rapidly decays in the neutral ground state S$_0$ by emitting a X$^{0}$ photon. For 2.3 V $<$ V $<$ 3.0 V, only T$_1$ is populated.  As T$_1$ is a long lived triplet state, a second tunneling event from the tip can take place, driving the QA molecule in the negatively charged doublet state D$_0^{-}$. From there, the tunneling of an electron from the HOMO to the sample can leave the molecule in the S$_1$ state, where radiative recombination to S$_0$ may occur. We note that, in contrast to the other many-body states discussed so far that are all directly observed either in conductance or fluorescence spectra, the presence of the D$_0^{-}$ state can, in principle, only be inferred from the current and voltage dependencies of the X$^{0}$ line. However, some information on this two-electron excitation mechanism can also be deduced from the difference between the X$^0$ line HRFM recorded at $V$ = 3.2 V and $V$ = 2.6 V (Fig.\,\ref{fig2}b-d). Whereas the first one can be well reproduced by considering a simple charge transfer from the tip to the LUMO of the molecule (reflecting the D$_0^{+} \rightarrow$ S$_1$ transition), the second involves the LUMO (D$_0^{+} \rightarrow$ T$_1$ transition) and the HOMO (T$_1 \rightarrow$ D$_0^{-}$ transition).

This can be accounted for in a theoretical rate-equation model where the input rates of all tip-mediated processes - tunneling and photon emission - are made as a function of the tip position. The tip-mediated tunneling rates (blue arrows in \Figref{fig3}d) are considered proportional to the relevant partial density of states\cite{hamann85,schulz2015many} and the plasmon-mediated photon emission (thick vertical arrows in \Figref{fig3}d) is included by evaluating the plasmon-exciton coupling strength\cite{roslawska2022prx}. The details of the rate-equation model and its parameterization are described in Supplemental Material. We solve the rate equations in the steady state for a given voltage $V$ at each position of the tip and thus extract the HRFM maps.
The calculated maps, shown in \Figref{fig2}d, are in an excellent agreement with the experiment, which demonstrates that, beyond the information related to the plasmon-exciton coupling, the detailed pattern of the HRFM can be used to identify the role of specific molecular many-body states in the STML excitation mechanism.

\begin{figure}[t]
\includegraphics[width=8.5cm]{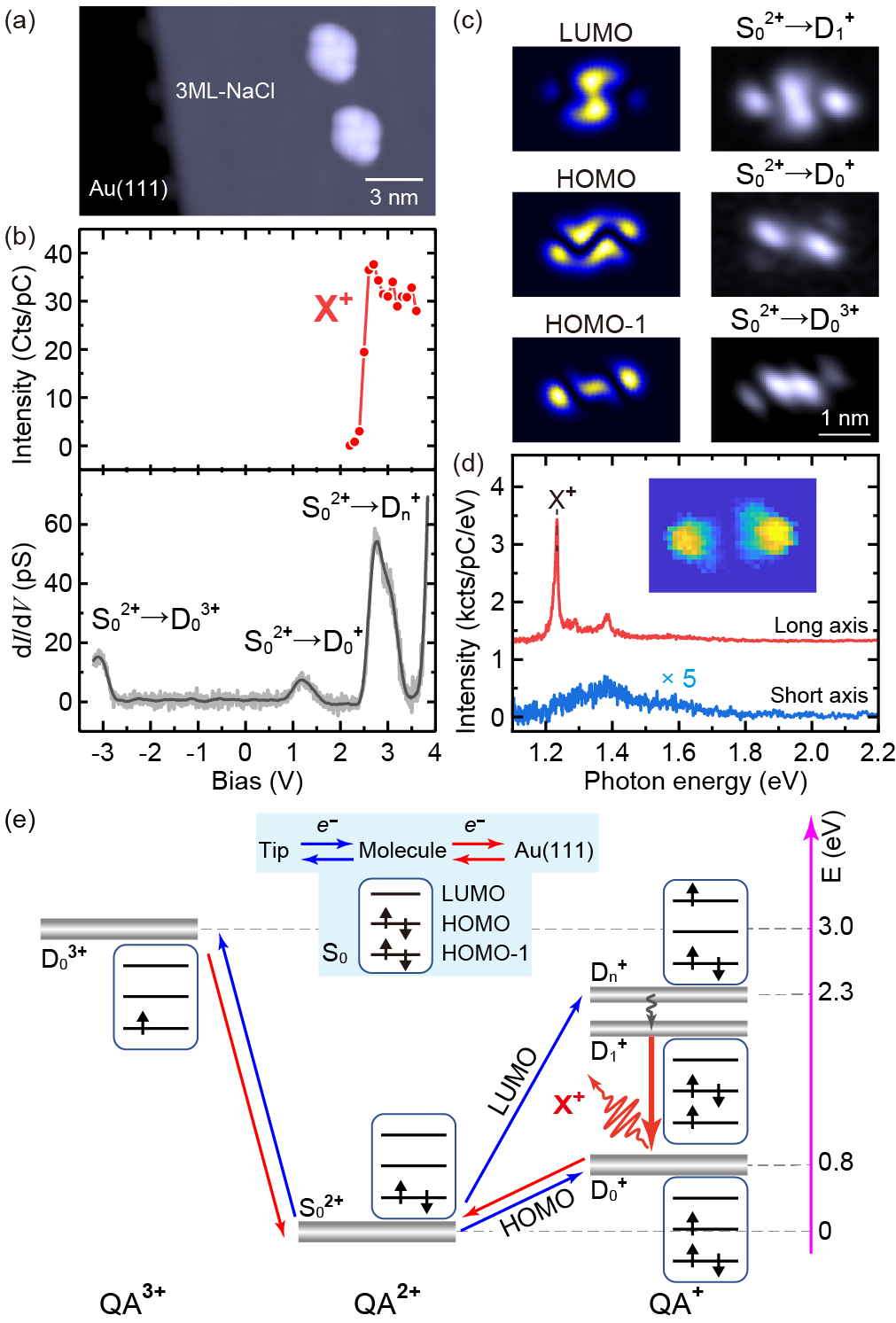}
\caption{\label{fig4} (a) STM image ($V$= 2.9 V, $I$ = 5 pA) of QA adsorbed on a 3ML-NaCl/Au(111). (b) Voltage dependency of the X$^+$ emission (top) and d$I$/d$V$ spectrum (bottom) recorded for QA/NaCl(3ML)/Au(111). (c) DFT  simulations of the partial density of states derived from the LUMO, HOMO and HOMO-1 of the neutral QA evaluated at a constant height above the molecule (left), and constant height d$I$/d$V$ maps (right) acquired at voltages corresponding to the resonances in the d$I$/d$V$ in (b). (d) STML spectra acquired ($V$ = 2.9 V, $I$ = 50 pA, $t$ = 180 s) for the STM tip located at the short and long axis of the molecule. Inset: HRFM  of the X$^+$ contribution  of QA on 3ML-NaCl/Au(111) (3.5 nm $\times$ 2.4 nm, 35 $\times$ 24 pixels, $V$ = 2.9 V, $t$ = 30 s per pixel.) (e) Sketch of the luminescence mechanisms for QA on 3ML-NaCl/Au(111).} 
\end{figure} 

To further confirm and extend our understanding of the excitation mechanism, we performed the same experiment with QA adsorbed on 3ML NaCl/Au(111), as shown in \Figref{fig4}a. Au(111) has a  work function higher by $\approx$ 0.8 eV than Ag(111), which causes a shift of the molecular states to higher energies, and eventually leads to an adsorbed molecule in a doubly charged singlet state S$_0^{2+}$. This unusual electronic configuration can be visualized in \Figref{fig4}b where the d$I$/d$V$ spectrum reveals three spectroscopic features assigned respectively to transitions to the triply charged doublet state D$_0^{3+}$ ($V$ = $-$3 V), the doublet state D$_0^{+}$ ($V$ = 1 V) and the excited doublet state D$_n^{+}$ ($V$ = 2.5 V) \cite{note}. 
STML spectra and HRFM reveal (\Figref{fig4}d) the characteristic features of the X$^{+}$ emission. The X$^{0}$ emission is absent, independently of the used voltage, current and tip position conditions. Indeed, as the molecule is initially in a doubly charged state, populating S$_1$, a mandatory step for the X$^{0}$ emission, would rely on an inefficient two-electron process and require a very high voltage to enable the D$_0^{+} \rightarrow$ S$_1$ transition. The X$^{+}$ voltage onset ($V \approx$ 2.3 V, see \Figref{fig4}b) correlates well with the  S$_0^{2+} \rightarrow$ D$_n^{+}$ transition, and suggests a non-radiative D$_n^{+} \rightarrow$ D$_1^{+}$ transition.  

On the basis of these observations one can therefore establish the many-body state diagram of \Figref{fig4}e. This diagram easily explains that the constant height d$I$/d$V$ images (\Figref{fig4}c) show patterns corresponding (i) to the LUMO  for the resonance corresponding to the S$_0^{2+} \rightarrow$ D$_n^{+}$ transition, (ii) to the HOMO for the resonance corresponding to the S$_0^{2+} \rightarrow$ D$_0^{+}$, and (iii) to the HOMO-1 for the resonance corresponding to the S$_0^{2+} \rightarrow$ D$_0^{3+}$ transition. Note that, in this last case, the molecule is transiently driven in an unexpected triply charged state. 
Eventually, we want to stress that our representation could be further improved by accounting for molecular vibrations and substrate reorganisations that are known to play an important role in the energy of the allowed electronic transitions \cite{fatayer2018reorganization, hernangomez2020}. 
 
Overall, we report on STML measurements of a model system for charged molecules. Our study  reveals that the emission line appearing strongly red-shifted with respect to the neutral fluorescence in this case should be assigned to trionic emission. Our paper also provides a very consistent picture based on current, voltage and spacial dependencies of STML signals as well as on conductance spectra and maps for the same molecule on two different substrates. This shows that a molecule can be populated in four different charged states and three different spin multiplicities during a single voltage sweep. This also allows establishing a solid many-body model explaining in details the transport and light emission mechanisms, which can readily be applied to other STML experiments dealing with decoupled molecules or more complex emitters.

We thank Niklas Humberg and Moritz Sokolowski for drawing our attention to the quinacridone molecule and for providing it. Anna Ros{\l}awska, Katharina Kaiser and Kirill Vasilev are also acknowledge for fruitful discussions. We are grateful to Virginie Speisser and Michelangelo Romeo for technical support. This project has received funding from the European Research Council (ERC) under the European Union's Horizon 2020 research and innovation program (grant agreement No 771850). The International Center for Frontier Research in Chemistry (FRC) is acknowledged for financial support. This work has been supported by the University of Strasbourg's IdEx program and by the ``Investissements d'Avenir'' LabEx PALM (ANR-10-LABX-0039-PALM).

\bibliographystyle{prsty}
\bibliography{Reference}

\begin{thebibliography}{10}
\bibitem{note}
A direct transition to D$_1^{+}$ is unlikely as it would require a change of
  electronic configuration involving two different electrons and would not
  account for the LUMO pattern in the S$_0^{2+} \rightarrow$ D$_n^{+}$ image in
  \Figref{fig4}c.
\end{thebibliography}
\end{document}